\documentclass[prl,preprint,endfloats]{revtex4-1}

\usepackage[dvipdfmx]{graphicx}
\usepackage{bm}
\usepackage{pifont}
\usepackage{amsmath}
\usepackage{soul}
\usepackage{color}

\begin{document}


\title{Zoology of multiple-$\bm Q$ spin textures in a centrosymmetric tetragonal magnet with itinerant electrons}

\author{Nguyen Duy Khanh$^{1,2,*}$, Taro Nakajima$^{1,2}$, Satoru Hayami$^{3,5}$, Shang Gao$^{1,10}$, Yuichi Yamasaki$^{4,5}$, Hajime Sagayama$^{6}$, Hironori Nakao$^{6}$, Rina Takagi$^{3,5,8}$, Yukitoshi Motome$^{3}$, Yoshinori Tokura$^{1,3,9}$, Taka-hisa Arima$^{1,7}$, Shinichiro Seki$^{1,3,5,8,*}$}

\affiliation{$^1$ RIKEN Center for Emergent Matter Science (CEMS), Wako, Japan}
\affiliation{$^2$ Institute for Solid State Physics (ISSP), University of Tokyo, Kashiwa, Japan}
\affiliation{$^3$ Department of Applied Physics, The University of Tokyo, Tokyo, Japan}
\affiliation{$^4$ Research and Services Division of Materials Data and Integrated System (MaDIS), National Institute for Materials Science (NIMS), Tsukuba, Japan}
\affiliation{$^5$ PRESTO, Japan Science and Technology Agency (JST), Kawaguchi, Japan}
\affiliation{$^6$ Institute of Materials Structure Science, High Energy Accelerator Research Organization, Tsukuba, Ibaraki, Japan}
\affiliation{$^7$ Department of Advanced Materials Science, The University of Tokyo, Kashiwa, Japan}
\affiliation{$^8$ Institute of Engineering Innovation, The University of Tokyo, Tokyo, Japan}
\affiliation{$^9$ Tokyo College, The University of Tokyo, Tokyo, Japan}
\affiliation{$^{10}$Current affiliation: Materials Science $\&$ Technology Division, Oak Ridge National Laboratory, Oak Ridge, TN, USA and Neutron Science Division, Oak Ridge National Laboratory, Oak Ridge, TN, USA}

\email{khanh.nguyen@riken.jp (N.D.K.); seki@ap.t.u-tokyo.ac.jp (S.S.)}


\begin{abstract}

{\bf Magnetic skyrmion is a topologically stable particle-like swirling spin texture potentially suitable for high-density information bit, which was first observed in noncentrosymmetric magnets with Dzyaloshinskii-Moriya interaction. Recently, nanometric skyrmion has also been discovered in centrosymmetric rare-earth compounds, and the identification of their skyrmion formation mechanism and further search of nontrivial spin textures are highly demanded. Here, we have exhaustively studied magnetic structures in a prototypical skyrmion-hosting centrosymmetric tetragonal magnet GdRu$_2$Si$_2$, by performing the resonant X-ray scattering experiments. We identified a rich variety of double-$\bf Q$ magnetic structures, including the antiferroic order of meron(half-skyrmion)/anti-meron-like textures with fractional local topological charges. The observed intricate magnetic phase diagram has been successfully reproduced by the theoretical framework considering the four-spin interaction mediated by itinerant electrons and magnetic anisotropy. The present results will contribute to the better understanding of the novel skyrmion formation mechanism in this centrosymmetric rare-earth compound, and suggest that itinerant electrons can ubiquitously host a variety of unique multiple-$\bf Q$ spin orders in a simple crystal lattice system.}

\end{abstract}

\maketitle

Multiple-$\bf Q$ spin order, i.e. two- or three-dimensionally modulated periodic structure characterized by a multiple number of coexisting modulation vectors $\bf Q$, has recently attracted attention as the source of rich emergent phenomena\cite{SkXReviewFertTwo, SkXReviewTokura, SkXTheoryFirst}. When the local magnetic moment ${\bf m} ({\bf r})$ takes a noncoplanar configuration, multiple-$\bf Q$ orders are often endowed with a nontrivial topology. For example, the superposition of multiple helical modulations can be regarded as the periodic array of skyrmions (Figure 1d), which appears as a stable particle-like object with a swirling ${\bf m}({\bf r})$ texture characterized by a nonzero integer topological charge $N_{\rm sk}=\int n_{\rm sk} dx dy$\cite{MnSi, TEMFeCoSi}. Here, the topological charge density $n_{\rm sk} ({\bf r})$ is defined as
\begin{equation}
n_{\rm sk} = \frac{1}{4\pi} {\bf n} \cdot \left(\frac{\partial {\bf n}}{\partial x} \times \frac{\partial {\bf n}}{\partial y} \right )
\label{skdensity}
\end{equation}
with ${\bf n}({\bf r})={\bf m}({\bf r})/|{\bf m}({\bf r})|$, and $N_{\rm sk}$ reflects how many times ${\bf n}$ wraps a unit sphere. Magnetic skyrmions are now intensively studied as a potential candidate of high-density information bit\cite{SkXReviewFertTwo} or building block of natural magnonic crystal \cite{magnonic} that is potentially suitable for magnetic data processing, and further explorations of nontrivial multiple-$\bf Q$ orders with exotic origin are highly demanded.

Previously, the skyrmion lattice (SkL) state with multiple-$\bf Q$ character was mostly observed in noncentrosymmetric magnets with the relativistic Dzyaloshinskii-Moriya (DM) interaction\cite{MnSi, TEMFeCoSi, Cu2OSeO3_Seki, CoZnMn_First, GaV4S8, Heusler}. In this case, the magnetic phase diagram is rather simple, where the single-$\bf Q$ helical phase is realized in zero magnetic field and the multiple-$\bf Q$ skyrmion lattice phase is stabilized by the application of magnetic field $\bf H$. On the other hand, the skyrmion formation has recently been discovered in centrosymmetric systems, such as Gd$_2$PdSi$_3$\cite{Gd2PdSi3}, Gd$_3$Ru$_4$Al$_{12}$\cite{GdKagome}, and GdRu$_2$Si$_2$\cite{GdRu2Si2}. Notably, these centrosymmetric Gd-compounds commonly host an extremely small skyrmion diameter ($\sim 2$ nm), which is one order of magnitude smaller than conventional DM-based noncentrosymmetric compounds. For such systems, several alternative skyrmion formation mechanisms have been proposed\cite{Frustration1, Frustration2, AkagiJPSJ, AkagiPRL, HayamiModel1, HayamiModel3, BatistaModel1, GdRu2Si2_STM, Batista2, utesov}, while their experimental verification has rarely been achieved.

In this work, we have exhaustively studied magnetic structures for a prototypical skyrmion-hosting centrosymmetric tetragonal magnet GdRu$_2$Si$_2$, by performing the resonant X-ray scattering (RXS) experiments. We identified a rich variety of double-$\bf Q$ magnetic structures, including the antiferroic order of meron/anti-meron-like textures with fractional local topological charge. The observed intricate magnetic phase diagram is successfully reproduced by the theoretical framework considering the four-spin interactions mediated by itinerant electrons and magnetic anisotropy. The present results will contribute to the better understanding of the novel skyrmion formation mechanism in this centrosymmetric rare-earth compound, and demonstrate that a variety of unique multiple-$\bf Q$ spin orders can be derived from a simple crystal lattice system with itinerant electrons.

Our target material GdRu$_2$Si$_2$ crystallizes into a centrosymmetric tetragonal structure with space group $I4/mmm$\cite{RRuSi2}, which consists of alternative stacking of the Gd square lattice and Ru$_2$Si$_2$ layers along the [001] axis (Figure 1a). Magnetism is governed by Gd$^{3+}$ ion with rather isotropic magnetic moment ($S = 7/2$, $L = 0$). Below the magnetic ordering temperature $T_N$ = 46 K, the appearance of incommensurate magnetic order with the modulation vector {\bf Q} = \{0.22, 0, 0\} has been reported\cite{GdRu2Si2}.

In Figure 1a and 1b, the $H-T$ (temperature) magnetic phase diagrams of GdRu$_2$Si$_2$ for $\textbf{H} \parallel [001]$ and $\textbf{H} \parallel [100]$ are summarized. The phase boundaries are determined by the magnetization measurements\cite{GdRu2Si2_PhaseDiagram}, where $\textbf H$-induced multiple-step metamagnetic transitions are observed for both $\textbf{H}$ directions (See Supplementary Note I and Supplementary Figure S2a). The phase II for $\textbf{H} \parallel [001]$ has been identified as the square skyrmion lattice state, i.e., the double-$\bf Q$ state described by the super-position of two orthogonally modulated screw structures\cite{GdRu2Si2} and a uniform magnetization (Figure 1d). Note that the recent scanning tunneling microscopy (STM) experiment \cite{GdRu2Si2_STM} revealed the two-dimensionally modulated (i.e. double-${\bf Q}$) electron density distribution in Phase II, III and possibly I, while this technique is rather sensitive to charge degree of freedom and the corresponding spin texture in the latter phases has not been confirmed directly. In addition, previous experiments mostly focused on $\textbf{H} \parallel [001]$, and the magnetic structures under $\textbf{H} \perp [001]$ remain unexplored.

To eluciadate the detailed magnetic structure in each phase, magnetic resonant X-ray scattering (RXS) experiments in resonance with the Gd L$_2$ edge at 5K were performed. First, we examined the development of the magnetic modulation vector ${\bf Q}$ for $\textbf{H} \parallel [100]$, by exploring the magnetic satellite reflections around a fundamental Bragg peak indexed as $(0, 4, 0) \pm {\bf Q}$ (Figure 2). Figure 2a and 2b indicate the line-scan profile along ($\delta$, 4, 0) and (0, 4-$\tau$, 0), which allows the identification of the magnetic modulation vectors  ${\bf Q}_1$ = ($q$, 0, 0) and ${\bf Q}_2$ = (0, $q$, 0) that are parallel and perpendicular to the external magnetic field $\bf H$, respectively. Here, the sample was initially cooled at $\mu_0H=0$ T, and the data points represented by open symbols were obtained in a field-increasing process from 0 T to 5 T. The ones represented by closed symbols were obtained in the subsequent field-decreasing process from 5 T to 0 T.  Magnetic-field dependence of the wavenumber $q$ and integrated intensity for these magnetic satellite reflections are also plotted in Figure 2c and 2d. After the initial zero-field cooling (i.e. the phase I), the magnetic satellite reflections are found for both $[100]$ and $[010]$ directions. By applying $\textbf{H} \parallel [100]$, the phase IV is stabilized for 2 T $< \mu_0H <$ 4 T, where the magnetic satellite reflection is observed only parallel to $\bf H$. This result indicates that the phase IV is the single-$\bf Q$ state with ${\bf Q}_1 \parallel {\bf H}$. Above 4 T (i.e. phase III'), the magnetic satellite reflection reappears along the both directions that are parallel and perpendicular to $\bf H$, which suggests that the phase III' is a double-$\bf Q$ state. When the $H$-value is reduced from the phase III', only ${\bf Q}_1 \parallel {\bf H}$ survives in the phase IV, but then both ${\bf Q}_1 \parallel {\bf H}$ and ${\bf Q}_2 \perp {\bf H}$ reappear in phase I. Here, the phase I is characterized by the anisotropic magnetic satellite reflections with ${\bf Q}_1 = (0.219, 0, 0)$ and ${\bf Q}_2 = (0, 0.224, 0)$. This indicates that the phase I is an anisotropic double-$\bf Q$ state which breaks the four-fold-symmetry. It leads to the appearance of two kinds of domains with opposite anisotropy in the phase I, and the application of $\textbf{H} \parallel [100]$ selects one of the magnetic domains. As detailed in Supplementary Note V, the phase V appearing for larger $H$-value has been identified as a single-$\bf Q$ state with ${\bf Q}_2 \perp {\bf H}$. In case of $\textbf{H} \parallel [001]$, both ${\bf Q}_1$ and ${\bf Q}_2$ magnetic satellite peaks are always observed in the phase I, II and III as discussed in Supplementary Note VI. 

Next, to understand the detailed relationship between the magnetic phases stabilized for $\textbf{H} \parallel [100]$ and $\textbf{H} \parallel [001]$, $M$-$H$ profiles have been measured for various directions of $\bf H$ rotated within the (010) plane (Supplementary Figure S2a). On the basis of these magnetization data, magnetic phase diagram as a function of $H_{[100]}$ and $H_{[001]}$ (i.e. the [100] and [001] components of $\bf H$) at 5 K is summarized in Figure 3j (See Supplementary Note II for the detail). Here, the angle between the $\bf H$-direction and the [001] axis is defined as $\theta$, and the corresponding $\theta$-dependence of magnetization at 3 T and 7 T are plotted in Figure 3k and 3l, respectively. At 3 T, the $M$-$\theta$ profile exhibits a clear step-like anomaly at $\theta = 60^\circ$, which represents the transition between the phases III and IV (Figure 3k). In contrast, the $M-\theta$ profile at $\mu_0H$ = 7 T shows a smooth behavior without any anomaly (Figure 3l), demonstrating the continuous transformation between the phase III in $\textbf{H} \parallel [001]$ and the phase III' (double-$\bf Q$ state) in $\textbf{H} \parallel [100]$. This observation suggests that the phase III is also a double-$\bf Q$ magnetic state, which is consistent with the recent STM observation of square lattice manner of charge density modulation in the phase III\cite{GdRu2Si2_STM}. To summarize, we conclude that the phases I, II, III, and III' are double-$\bf Q$ states, and that the phases IV and V are single-$\bf Q$ states.

To investigate the detailed spin texture in each phase, we have performed the polarization analysis of the scattered X-ray in the RXS experiments. The schematic illustration of the experimental setup is shown in Figure 3c. The magnetic field \textbf{H} is applied perpendicular to the scattering plane. The incident X-ray is polarized parallel to the scattering plane ($\pi$-polarization). The scattered X-ray beam may consist of two polarization components parallel ($\pi '$) and perpendicular ($\sigma '$) to the scattering plane, and their intensities ($I_{\pi - \pi '}$ and $I_{\pi - \sigma '}$) are measured separately. When the magnetic structure ${\bf m} ({\bf r})$ includes the modulated spin component ${\bf m}_{\bf Q} \exp (i{\bf Q} \cdot {\bf r})+c.c.$ (with ${\bf m}_{\bf Q}$ being a complex vector), the magnetic scattering intensity $I$ is given by\cite{RXS_Rule}
\begin{equation}
I \propto | ({\bf e}_i \times {\bf e}_f ) \cdot {\bf m}_{\bf Q} |^2.
\end{equation}
Here, ${\bf e}_i$ and ${\bf e}_f$ are unit vectors representing the polarization of incident and scattered X-ray beams, respectively. On the basis of the above relationship, we can separately evaluate the amplitude of each modulated spin component.

In the following, we discuss the case for the double-$\bf Q$ structure in phase III' at $\mu_0H =$ 5 T applied along the [100] direction, as an example. First, the magnetic satellite peaks at $(0, 4, 0) \pm {\bf Q}$ are investigated with the setup shown in Figure 3a. In this configuration, the propagation vector ${\bf k}_i$ of incident X-ray is almost parallel to the $[0{\bar 1}1]$ direction. Eq. (2) suggests that $I_{\pi - \pi'}$ and $I_{\pi - \sigma'}$ mainly reflect the [100] and $[0\bar{1}1]$ components of ${\bf m}_{\bf Q}$, respectively. Figure 3d and 3g indicate the line scan profiles of the $I_{\pi - \pi'}$ and $I_{\pi - \sigma'}$  intensities for ${\bf Q}_1$ = ($q$, 0, 0) and ${\bf Q}_2$ = (0, $q$, 0), respectively. In both cases, $I_{\pi - \sigma'}$ is clearly observed, while $I_{\pi - \pi'}$  is almost negligible. This suggests that ${\bf m}_{{\bf Q}_1}$ and ${\bf m}_{{\bf Q}_2}$ contain [010] or [001] component, but do not have the [100] component. We have further performed a similar polarization analysis for the magnetic reflections at $(0, 4, 4) \pm {\bf Q}$ with the setup shown in Figure 3b, where ${\bf k}_i$ is almost parallel to the $[0\bar{1}0]$ direction. In this case, $I_{\pi - \pi'}$ and $I_{\pi - \sigma'}$ mainly reflect the [100] and [010] components of ${\bf m}_{\bf Q}$, respectively. Figure 3e indicates the line scan profiles of $I_{\pi - \pi'}$ and $I_{\pi - \sigma'}$ for ${\bf Q}_1$. The observed presence of $I_{\pi - \sigma'}$ and absence of $I_{\pi - \pi'}$ suggest that ${\bf m}_{{\bf Q}_1}$ mainly consists of the [010] component. The corresponding line scan profile for ${\bf Q}_2$ is also shown in Figure 3h, where the absence of $I_{\pi - \pi'}$ and $I_{\pi - \sigma'}$ intensities demonstrates that ${\bf m}_{{\bf Q}_2}$ contains neither [010] nor [100] component. Since the aforementioned results for $(0, 4, 0) \pm {\bf Q}_2$ (Figure 3g) suggest that the projection of ${\bf m}_{{\bf Q}_2}$ along the $[0\bar{1}1]$ direction is not zero, ${\bf m}_{{\bf Q}_2}$ should possess the [001] component. On the basis of the above results, we conclude that the spin texture in the phase III' for $\textbf{H} \parallel [100]$ can be approximately described as
\begin{equation}
{\bf m}({\bf r}) \propto [(0,m_b,0) \exp(i{\bf Q}_1 \cdot {\bf r})+(0,0,m_c) \exp(i {\bf Q}_2 \cdot {\bf r}) +c.c. ] + M_0 (1,0,0)
\label{PhaseIII'}
\end{equation}
with $M_0$ representing the $\bf H$-induced uniform magnetization component. The resultant anisotropic double-$\bf Q$ spin texture for phase III' is the superposition of two orthogonally modulated sinusoidal spin components (Figure 3f and 3i) as shown in Figure 1f. 

By performing similar RXS measurements with the polarization analysis, each ${\bf m}_{\bf Q}$ component has been evaluated for all the magnetic phases (i.e. phase I - V). The experimentally identified spin textures are summarized in Figure 1c-h. (See Supplementary Note III, IV, V and VI for the detail). The phase I, which has been identified as an anisotropic double-$\bf Q$ state from the aforementioned discussion for Figure 2, turns out to possess the spin texture described by 
\begin{equation}
{\bf m}({\bf r}) \propto [(0,m_b,im_c) \exp(i{\bf Q}_1 \cdot {\bf r})+(m_a,0,0) \exp (i{\bf Q}_2 \cdot {\bf r})] + c.c.,
\end{equation}
i.e. the superposition of screw and sinusoidal spin modulations with ${\bf Q}_1 = (0.219, 0, 0)$ and ${\bf Q}_2 = (0, 0.224, 0)$, respectively (Figure 1c). On the other hand, the phase III induced by $\textbf{H} \parallel [001]$ is the double-$\bf Q$ state given by 
\begin{equation}
{\bf m}({\bf r}) \propto [(0,m_b,0) \exp(i{\bf Q}_1 \cdot {\bf r})+(m_a,0,0) \exp (i{\bf Q}_2 \cdot {\bf r}) + c.c.] + M_0 (0,0,1),
\label{PhaseIII}
\end{equation}
which represents the square vortex lattice state described by the superposition of two sinusoidally modulated spin components (Figure 1e). The phases IV and V induced by $\textbf{H} \parallel [100]$ are the single-$\bf Q$ states described by
\begin{equation}
{\bf m}({\bf r}) \propto [(0,m_b,im_c)  \exp(i{\bf Q}_1 \cdot {\bf r}) + c.c.] + M_0 (1,0,0)
\end{equation}
and
\begin{equation}
{\bf m}({\bf r}) \propto [(0,0,m_c) \exp(i{\bf Q}_2 \cdot {\bf r}) + c.c.] + M_0 (1,0,0),
\end{equation}
representing the screw spin order with ${\bf Q}_1 \parallel {\bf H}$ (Figure 1g) and the sinusoidally modulated fan-like spin order with ${\bf Q}_2 \perp {\bf H}$ (Figure 1h), respectively. Notably, the experimentally identified ${\bf m}_{\bf Q}$ in the $H$-induced phases III, III', IV and V are always normal to the $\bf H$-direction (Figure 1e-h). In general, the magnetic moments in the antiferromagnetic and helimagnetic states favor to be aligned perpendicular to the $\bf H$-direction, and the observed ${\bf m}_{\bf Q} \perp {\bf H}$ relationship is reasonable in this context. At 7 T, the rotation of magnetic field direction from the [001] axis (phase III represented by Eq. (\ref{PhaseIII})) to the [100] axis (phase III' represented by Eq. (\ref{PhaseIII'})) leads to the continuous rotation of ${\bf m}_{{\bf Q}_2}$ within the (010) plane keeping the ${\bf m}_{{\bf Q}_2} \perp {\bf H}$ relationship, which well explains the observed smooth $M$-$\theta$ profile in Figure 3l.

Next, we discuss the microscopic mechanism to realize such an intricate magnetic phase diagram with a rich variety of double-$\bf Q$ orders. In the following, we demonstrate that the observed phase diagram and spin textures can be well reproduced by the effective Hamiltonian given by
\begin{equation}
\mathcal{H}=2N\sum_{\nu=1,2} \left [ -J \left ( \sum_{\alpha=x,y,z} \Gamma_{{\bf Q}_\nu}^{\alpha} m_{{\bf Q}_\nu}^{\alpha} m_{-{\bf Q}_\nu}^{\alpha} \right )  + K \left ( \sum_{\alpha=x,y,z} \Gamma_{{\bf Q}_\nu}^{\alpha} m_{{\bf Q}_\nu}^{\alpha} m_{-{\bf Q}_\nu}^{\alpha} \right )^2 \right ] - \sum_{i} {\bf H} \cdot {\bf m}_i,
\label{HayamiHamiltonian}
\end{equation}
which is derived from a Kondo lattice model consisting of localized magnetic moments and itinerant electrons. This theoretical framework has originally been proposed in Refs. \cite{HayamiModel1, HayamiModel2}, where a set of wavenumbers ${\bf Q}_1=(q, 0, 0)$ and ${\bf Q}_2=(0, q, 0)$ are assumed to take peaks in the bare susceptibility under the tetragonal lattice symmetry. $J (\equiv 1)$ represents the amplitude of Ruderman-Kittel-Kasuya-Yosida (RKKY) interaction stabilizing the magnetic modulation with the wavevector ${\bf Q}_\nu$, and $K$ represents the amplitude of four-spin interaction favoring multiple-$Q$ orders\cite{HayamiModel1, HayamiModel2, Blugel}. $\Gamma_{{\bf Q}_1}=(\gamma_x, \gamma_y, \gamma_z)$ and $\Gamma_{{\bf Q}_2}=(\gamma_y, \gamma_x, \gamma_z)$ describe the anisotropy of associated interactions originating from the spin-orbit coupling. The last term represents the Zeeman coupling in an external magnetic field {\bf H}, and ${\bf m}_i$ represents the local magnetic moment at the position $i$  (See Methods section for the detailed parameters and procedures for the theoretical simulation.)

Figure 4a shows the magnetization profile for $\textbf{H} \parallel [001]$ theoretically calculated by Eq. (\ref{HayamiHamiltonian}) with $K = 0.3$. It shows a series of step-like anomalies, in good agreement with the experimental results (Supplementary Figure S2a). The corresponding $H$-dependence of individual $(m^\alpha_{{\bf Q}_\nu})^2$ component and the simulated spin texture at selected $H$-values are shown in Figure 4b and 4g-i, respectively. We have found that the experimentally identified spin textures for phases I (anisotropic double-$\bf Q$ state (Figure 1c)), II (double-$\bf Q$ square skyrmion lattice state (Figure 1d)) and III (double-$\bf Q$ square vortex lattice state (Figure 1e)), as well as the $\bf H$-induced transitions among these phases (i.e. I $\rightarrow$ II $\rightarrow$ III $\rightarrow$ FM (Figure 1a)), are successfully reproduced in the present simulation. The same behaviors are also confirmed for a wide range of $K$ parameter of $0.1 < K \lesssim 0.5$.

Similar theoretical calculations have been also performed for $\textbf{H} \parallel [100]$, and the obtained $H$-dependence of $\bf M$ and $(m_{{\bf Q}_\nu}^\alpha)^2$, as well as the associated spin textures at selected $H$-values, are summarized in Figure 4c-f and 4j-l. The simulation well reproduces the experimentally identified spin textures for the phases IV (single-$\bf Q$ screw state (Figure 1g)), III' (anisotropic double-$\bf Q$ state (Figure 1f)) and V (single-$\bf Q$ fan state (Figure 1h)), while the manner of $\bf H$-induced phase transition depends on the magnitude of $K$. The present calculation predicts the successive transitions in order of I $\rightarrow$ IV $\rightarrow$ V $\rightarrow$ FM (Figure 4c) and I $\rightarrow$ III' $\rightarrow$ V $\rightarrow$ FM (Figure 4e) for $K = 0.2$ and $K=0.3$, respectively, which suggests that the free energies for phases III' and IV are almost degenerated. In the experiment, the transition in order of I $\rightarrow$ IV $\rightarrow$ III' $\rightarrow$ V $\rightarrow$ FM is observed (Figure 1b) in accord with theoretical prediction, implying that the fine tuning of $K$ value and/or magnetic anisotropy would allow the full reproduction of the experimental phase transition process for $\textbf{H} \parallel [100]$. The above results suggest that Eq. (\ref{HayamiHamiltonian}) well captures the physics behind the skyrmion formation and the related multiple-$\bf Q$ magnetism in this system.

To understand the topological aspect of each spin texture, their corresponding spatial distribution of topological charge density $n_{\rm sk} ({\bf r})$ are plotted in Figure 4m-r. The double-$\bf Q$ phases I, II and III are characterized by the common texture of in-plane spin component (Figure 4g-i), which possesses two vortices and two anti-vortices in the magnetic unit cell. By further considering the out-of-plane spin component, they can be considered as a periodic lattice of meron and anti-meron-like texture, where magnetic moments in meron and anti-meron wrap a half of unit sphere and are characterized by $N_{\rm sk}=-1/2$ and $N_{\rm sk}=+1/2$, respectively \cite{Meron1,Meron3, Meron2}. According to Eq. (1), the sign of $N_{\rm sk}$ is affected by the orientation of out-of-plane spin component at their core positions and the swirling manner of in-plane spin component in the surrounding area. For example, the magnetic unit cell of the phase II consists of a core-down vortex (i.e. a meron with $N_{\rm sk}=-1/2$), a core-up vortex (i.e. an anti-meron with $N_{\rm sk}=+1/2$) and two core-up anti-vortices (i.e. two merons), which can be viewed as the square lattice of skyrmion ($N_{\rm sk}=-1$) (Figure 4n). On the other hand, the phases I and III are characterized by the different orientation of core magnetization, and can be considered as the antiferroic order of meron/anti-meron-like textures into the stripe and checkerboard patterns, respectively (Figure 4m and 4o). Likewise, the double-$\bf Q$ phase III' can be also interpreted as the checkerboard antiferroic order of meron/anti-meron-like textures (Figure 4q). Here, the phases III and III' smoothly transform into each other by rotating magnetic field, keeping the ${\bf m}_{\bf Q} \perp {\bf H}$ relationship. Since $n_{\rm sk} ({\bf r})$ is invariant under the global rotation of magnetic moments according to Eq. (1), these two phases are characterized by the common spatial distribution of $n_{\rm sk}$ as shown in Figure 4o and 4q.

In general, conduction electrons interacting with such noncoplanar magnetic textures feel the local emergent magnetic field $b_{\rm em}^{z} ({\bf r}) = (h/e) n_{\rm sk} ({\bf r})$ due to the additional quantum Berry phase\cite{SkXReviewTokura, THE, EmergentEfield}. In case of GdRu$_2$Si$_2$ with magnetic modulation period $\lambda \sim 1.9$ nm\cite{GdRu2Si2}, the amplitude of $b_{\rm em}^{z} ({\bf r})$ should be in order of $\sim 100$ T. The above analysis suggests that the presently identified double-$\bf Q$ phases are characterized by the periodic modulation of giant emergent magnetic field $b_{\rm em}^{z} ({\bf r})$, which will lead to the exotic manner of local electron dynamics (such as meandering motion of electrons). Note that $n_{\rm sk} ({\bf r})$ is always negligible for the phase IV and V with single-$\bf Q$ character (Figure 4p,r), suggesting that the multiple-$\bf Q$ nature is essential for the appearance of local topological charge and associated emergent magnetic field. 

In Figure 4s-x, we plot theoretically calculated reciprocal-space distribution of $|{\bf m}_{\bf Q}|$ for each phase. They always exhibit peak structures corresponding to $\bf Q_1$ and $\bf Q_2$. On the other hand, the higher order peak at the $\bf Q_1+Q_2$ position can be identified only in phase II (square skyrmion lattice phase), but not in the other double-$\bf Q$ phases. Since the magnetic scattering intensity in the diffraction experiment is generally proportional to $|{\bf m}_{\bf Q}|^2$, the absence of $\bf Q_1+Q_2$ reflection in phases I and III reported in \cite{GdRu2Si2} would be reasonable.




In this study, we have identified a variety of double-$\bf Q$ spin textures (including the antiferroic order of meron/antimeron-like textures with fractional local topological charge) for a prototype skyrmion-hosting centrosymmetric tetragonal magnet GdRu$_2$Si$_2$. The observed good agreement between the experimental and theoretical phase diagrams suggests a potential mechanism for the skyrmion formation in this system, where the magnetic interactions mediated by itinerant electrons and the associated magnetic anisotropy play a significant role. Note that several alternative theoretical frameworks based on magnetic frustration has also been proposed recently, some of which may be effectively mapped into the present model and possibly reproduce the observed phase diagrams by tuning their parameters \cite{Batista2,utesov}. These novel theoretical frameworks predict the appearance of numerous multiple-$\bf Q$ orders consisting of various topological solitons such as merons, skyrmions and higher-order skyrmions with $N_{\rm sk}= \pm 1/2$, 1, and 2, respectively, depending on the symmetry of underlying crystal lattice and magnetic anisotropy \cite{HayamiModel1, HayamiModel2, HayamiModel3, BatistaModel1}. Previously, such multiple-${\bf Q}$ spin textures with fractional or higher-order topological charge have rarely been identified experimentally \cite{Meron1, Meron3}. The present results glimpse how rich variety of multiple-$\bf Q$ orders with unique topology/symmetry can be derived from a simple crystal lattice system with itinerant electrons, and highlight rare-earth intermetallics as a promising platform for the further search of exotic topological soliton ensembles with nontrivial functionality. The direct real-space observation of these textures and full theoretical reproduction of finite temperature phase diagram are the issue for the future study.

\begin{figure}
\begin{center}
\includegraphics*[width=12.5cm, keepaspectratio]{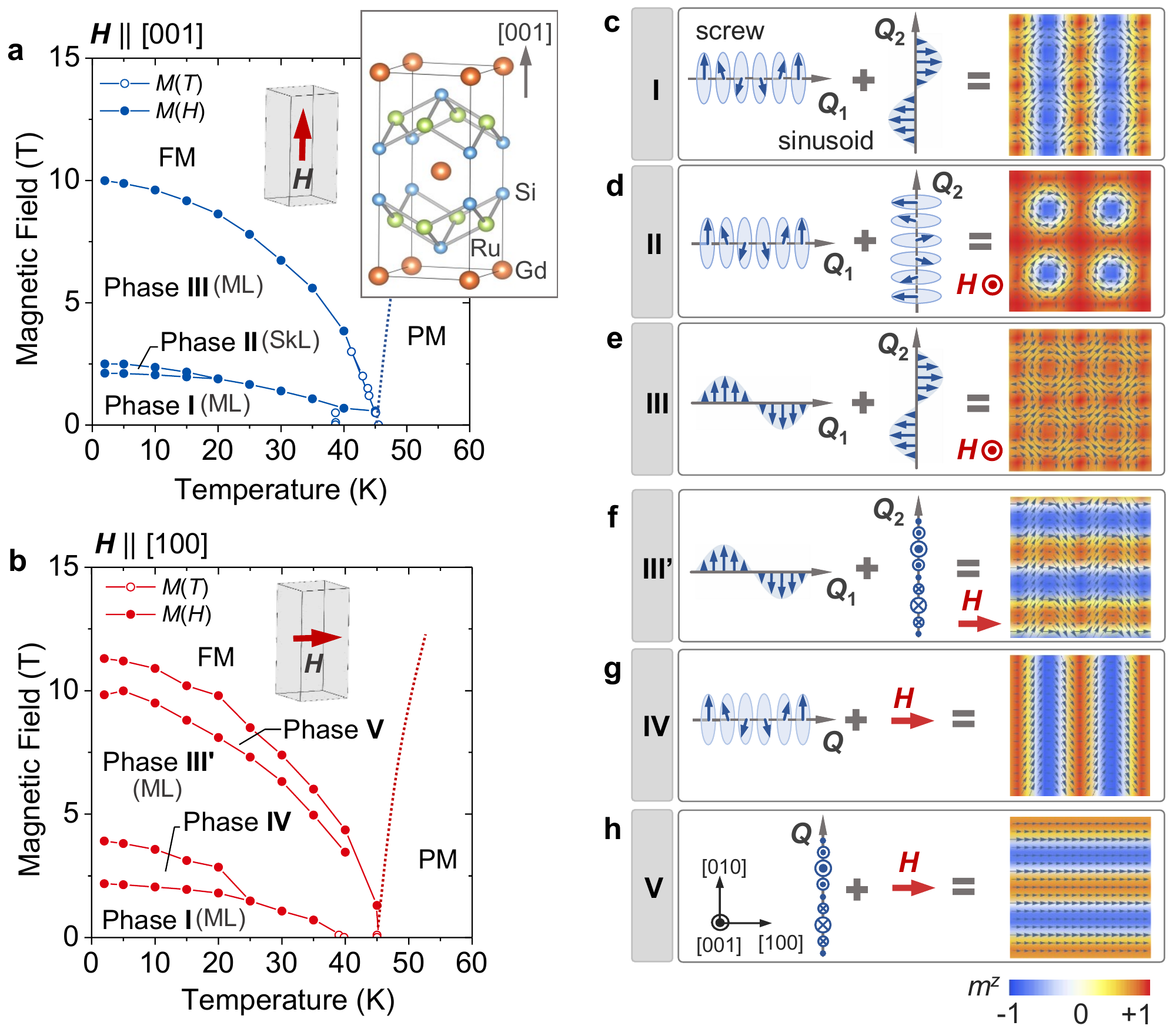}
\caption{{Magnetic phase diagrams and magnetic structures for GdRu$_2$Si$_2$.} a,b) $H$(magnetic field)-$T$(temperature) magnetic phase diagrams for $\textbf{H} \parallel [001]$ (a) and $\textbf{H} \parallel [100]$ (b) determined based on the magnetization measurements. SkL and ML denote the double-$\bf Q$ spin orders representing the square lattice of skyrmion texture and the antiferroic lattice of meron/anti-meron-like textures, respectively. The crystal structure of GdRu$_2$Si$_2$ is also shown in the inset. c-h) The schematic illustration of spin texture for the phase I, II, III, III', IV and V, identified through the resonant X-ray scattering experiments with polarization analysis. The presented spin textures are based on Eqs. (3)-(7), with their local amplitude of Gd magnetic moment kept constant. The phases I, II, III and III' are the double-$\bf Q$ states described by the superposition of two orthogonally modulated spin textures. The background color represents the amplitude of out-of-plane spin component $m^z$.}
\end{center}
\end{figure}

\begin{figure}
\begin{center}
\includegraphics*[width=17.0cm, keepaspectratio]{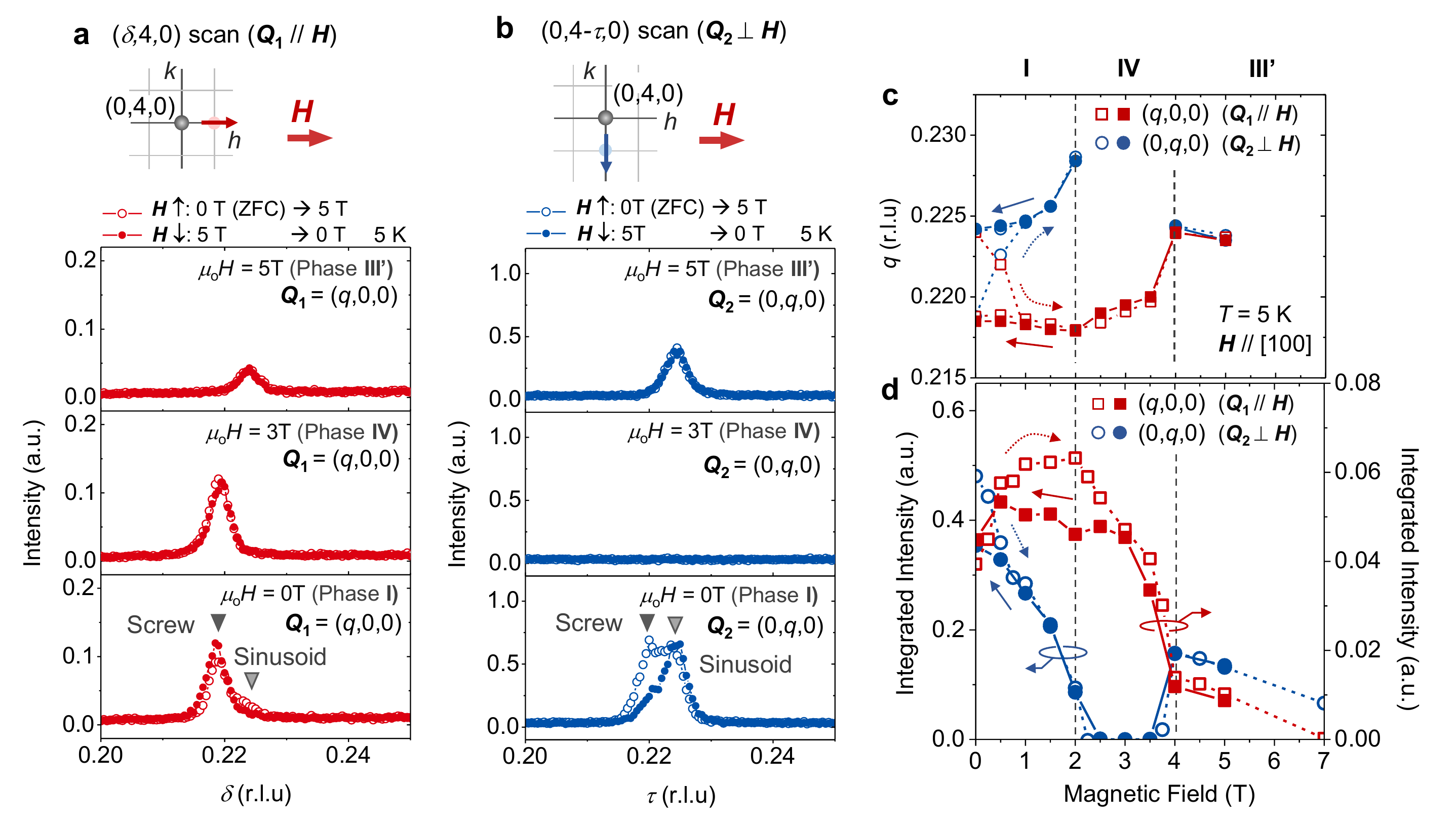}
\caption{{Magnetic field dependence of resonant X-ray scattering profiles for $\bf Q_1$ and $\bf Q_2$ magnetic satellite reflections under $\textbf{H} \parallel [100]$.} a) Line profiles for ($\delta$, 4, 0) scan measured with various amplitude of $\textbf{H} \parallel [100]$ at 5 K, which represents the ${\bf Q}_1 (\parallel {\bf H})$ magnetic satellite peak around the fundamental Bragg spot (0, 4, 0). The sample was initially cooled at $\mu_0H=0$, and the data points represented by open symbols are first measured in the field increasing process from 0 T to 5 T. Then, the one represented by closed symbols are measured in the field decreasing process from 5 T to 0 T. The schematic illustration of line-scan direction in the reciprocal space is shown in the upper panel. b) The corresponding data for (0, $4-\tau$, 0) scan, representing the ${\bf Q}_2 (\perp \bf H)$ magnetic satellite peak. c,d) Magnetic field dependence of wavenumber $q$ and integrated intensity for magnetic satellite reflections, obtained from the data sets as shown in (a) and (b) (See Supplementary Figure S3 for the detail). The red and blue symbols represent the data for ${\bf Q}_1$ $(\parallel {\bf H})$ and ${\bf Q}_2$ $(\perp {\bf H})$, and the open and closed symbols represent the one measured in the $H$-increasing and decreasing process, respectively.}
\end{center}
\end{figure}

\begin{figure}
\begin{center}
\includegraphics*[width=17cm, keepaspectratio]{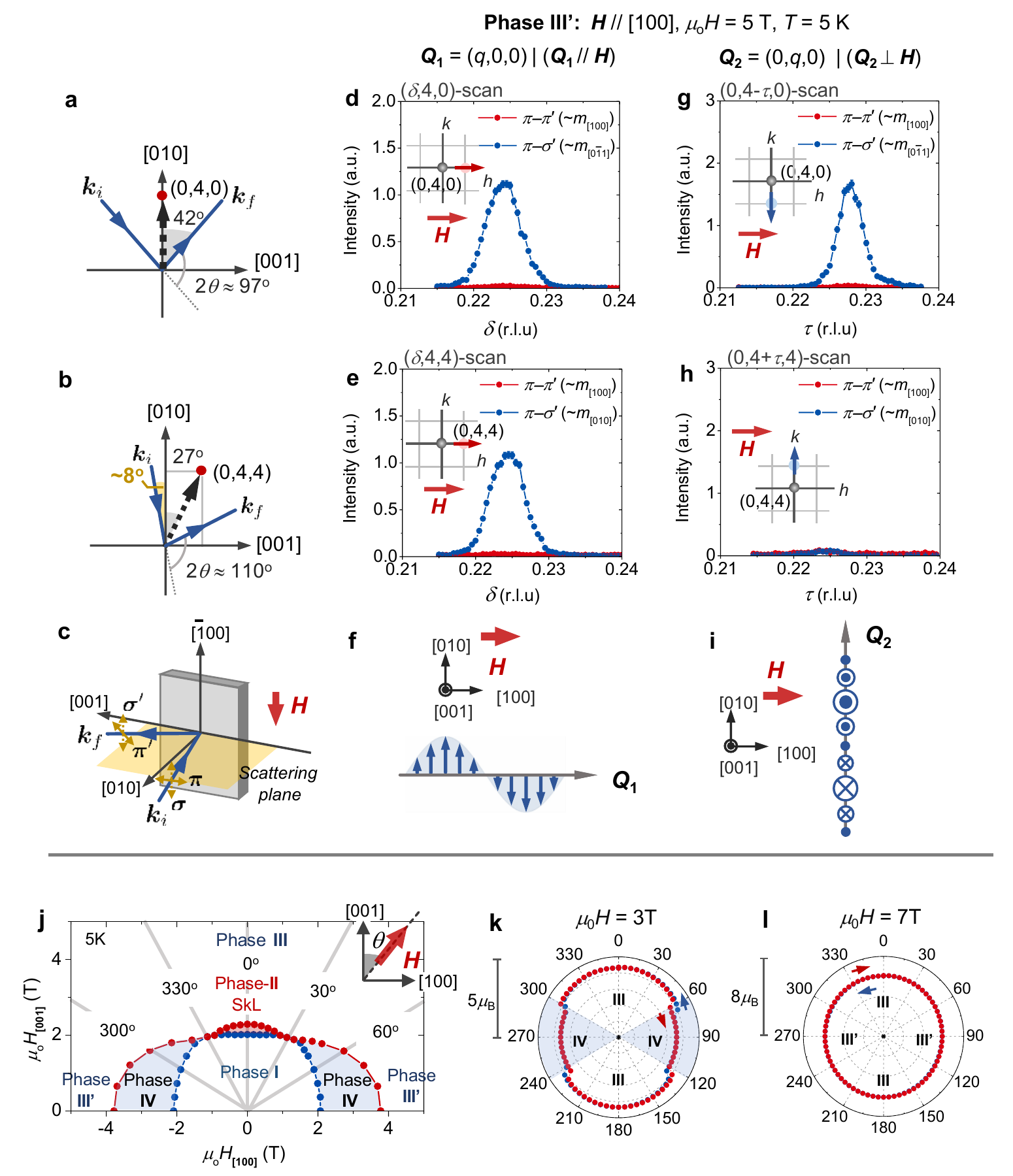}
\caption{{Polarization analysis of RXS profiles in the phase III', and magnetic phase diagram as a function of $H_{[100]}$ and $H_{[001]}$.} a, b) Experimental configuration of RXS measurement to investigate magnetic satellite peaks around the fundamental Bragg spots (0, 4, 0) and (0, 4, 4), respectively. Scattering plane is always perpendicular to the [100] axis. ${\bf k}_i$ and ${\bf k}_f$ are the propagation vectors of incident and scattered X-ray, respectively. In (a) and (b), ${\bf k}_i$ is almost parallel to the [0$\bar{1}$1] and [0$\bar{1}$0] direction, respectively. c) Schematic illustration of experimental setup for polarization analysis. $\pi$ ($\pi '$) and $\sigma$ ($\sigma '$) are the polarization direction of incident (scattered) X-ray. d, e) Line profiles for ($\delta$, 4, 0) and ($\delta$, 4, 4) scans for ${\bf Q}_1$ ($\parallel {\bf H}$) satellite reflection, measured at 5 K with $\mu_0H$ = 5 T applied along the [100] direction (phase III').g, h) The corresponding line profiles for (0, $4-\tau$, 0) and (0, $4+\tau$, 4) scans for ${\bf Q}_2$ ($\perp {\bf H}$) satellite reflection. f, i) Real-space illustration of modulated spin components that belong to each magnetic modulation vector ${\bf Q}_1$ and ${\bf Q}_2$. j) Magnetic phase diagram as a function of $H_{\rm [100]}$ and $H_{\rm [001]}$ (i.e. the [100] and [001] component of $\bf H$, respectively), determined from the $M$-$H$ profiles measured for various direction of $\bf H$ at 5 K (Supplementary Figure 2). The angle between the $\bf H$-direction and the [001] axis is defined as $\theta$. k, l) $\theta$-dependence of magnetization measured at $\mu_0H$  = 3 T ({\it K}) and 7 T ({\it L}).}
\end{center}
\end{figure}

\begin{figure}
\begin{center}
\includegraphics*[width=16.5cm, keepaspectratio]{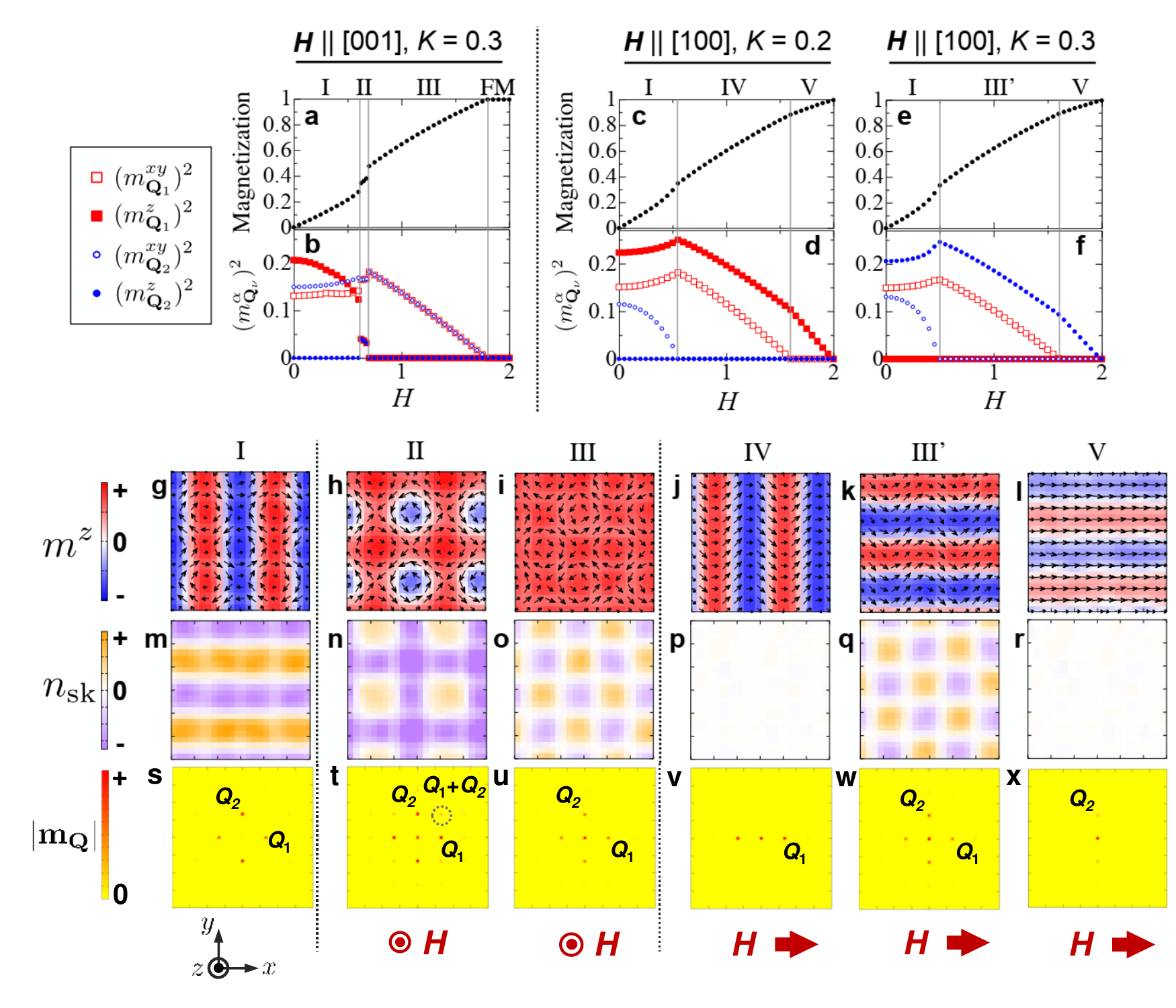}
\label{fig4}
\caption{{Numerical simulation of magnetic phase diagram and magnetic structures.} a, b) Magnetic field dependence of magnetization $\bf M$ and modulated spin component $(m^\alpha_{{\bf Q}_\nu})^2$, theoretically calculated by simulated annealing for $\textbf{H} \parallel [001]$ with Eq. (\ref{HayamiHamiltonian}) and $K=0.3$. Here, the uniform ferromagnetic (FM) state gives $M \simeq 1$, and we defined $(m^{xy}_{{\bf Q}_\nu})^2 \equiv (m^{x}_{{\bf Q}_\nu})^2+(m^{y}_{{\bf Q}_\nu})^2$. c-f) The corresponding results for $\textbf{H} \parallel [100]$ calculated with $K=0.2$ ((c) and (d)) and $K=0.3$ ((e) and (f)). g-x) The real space distribution of local magnetization ${\bf m}({\bf r})$ (g-l) and topological charge density $n_{\rm sk}({\bf r})$ (m-r) as well as (s-x) reciprocal-space distribution of $|{\bf m}_{\bf Q}|$ for the phase I, II, III, IV, III', and V obtained by simulated annealing. The background color represents the amplitude of $m^z$, $n_{\rm sk}$ and $|{\bf m}_{\bf Q}|$. The results for the single-$\bf Q$ state are obtained by averaging over 500 Monte Carlo samplings to reduce the effect of thermal fluctuations. }
\end{center}
\end{figure}

\section*{Methods} 

\subsection{Crystal Growth} 
Single crystals of GdRu$_2$Si$_2$ were grown by the optical floating zone method as reported in Ref. \cite{GdRu2Si2}. The phase purity of the sample was confirmed by the powder x-ray diffraction. The orientations of single crystals were determined using the back-reflection X-ray Laue photography method.

\subsection{Magnetic and electrical transport property measurements}

Magnetization measurements were performed using a superconducting quantum interference device magnetometer (MPMS, Quantum Design), which is accompanied with a rotation probe to characterize the angle dependence of magnetization. A single crystal of GdRu$_2$Si$_2$ was cut into a rectangular plate shape and silver paste was painted as electrodes for five terminal measurements to investigate electrical transport properties. Longitudinal resistivity $\rho_{xx}$ and transverse resistivity $\rho_{yx}$ were measured using the AC-transport option in a Physical Property Measurement System (PPMS, Quantum Design). The same sample was used for the measurements of magnetic and electrical transport properties to avoid the influence of the demagnetization effect, which may cause discrepancy in the phase diagram due to shape anisotropy.

\subsection{Resonant X-ray Scattering (RXS) measurement}

RXS experiments were carried out on BL-3A, Photon Factory, KEK, Japan. The measurements were performed using incident X-rays with a photon energy turned near the Gd $L_2$ absorption edge ($\sim$7.935 keV) by utilizing a Si (111) double-crystal monochromator. A single crystal of GdRu$_2$Si$_2$ with flat (010) plane was fixed on an Al plate using varnish and put into a vertical-field superconducting magnet. In this set up, the scattering plane (0, $K$, $L$) is perpendicular to the [100] axis. The incident X-ray was linearly polarized parallel to the scattering plane ($\pi$). To analyze the polarization of the scattered X-ray, the (006) reflection of a pyrolytic graphite (PG) plate was used. The $2\theta$ angle for the analyzer at Gd $L_2$ edge was $88^\circ$. The polarization components $\sigma'$ (perpendicular to the scattering plane) and $\pi'$ (parallel to the scattering plane) of scattered X-ray beams can be chosen by rotating the PG plate about the scattered beam.

\subsection{Theoretical calculation}

The results in Figure 4a-f were obtained by performing simulated annealing for the model in Eq.~(\ref{HayamiHamiltonian}). 
In the simulations, starting from a random spin configuration, the temperature was gradually reduced from high temperature to the final temperature $T=0.01$ with the rate $T_{n+1}=\alpha T_{n}$, where $T_n$ was the temperature in the $n$th step and $\alpha$ was set to be $0.99995$-$0.99999$.
The standard Metropolis local updates were performed in real space at each temperature and $10^5$-$10^6$ Monte Carlo sweeps were performed for thermalization and measurements at the final temperature. 
To determine the phase boundaries between different magnetic phases, we also performed the simulations from the spin patterns obtained at low temperatures.

The magnetic phases in Figure~4 were calculated for the model parameters $q=\pi/3$, $\gamma_x=0.855$, $\gamma_y=0.9$, $\gamma_z=1$, and $K=0.3$ [Figure~4a,b and 4e,f] and $K=0.2$ [Figure~4c and 4d] in the system size with $N=96^2$. 
In the case of ${\bf H} \parallel [001]$, three distinct phases were obtained in addition to the fully-polarized state at high fields; phase I for $0 \leq H \lesssim 0.6$, phase II for $0.6 \lesssim H \lesssim 0.7$, and phase III for $0.7 \lesssim H \lesssim 1.8$. 
On the other hand, three phases were stabilized in the case of ${\bf H} \parallel [100]$; phase I for $0 \leq  H \lesssim 0.55$, phase IV for $0.55 \lesssim H \lesssim 1.6$, and phase V for $1.6 \lesssim H \lesssim 2$ at $K=0.2$ and phase I for $H \lesssim 0.5$, phase III' for $0.5 \lesssim H \lesssim 1.6$, and phase V for $1.6 \lesssim H \lesssim 2$ at $K=0.3$. Note that we obtained the similar low-temperature phase sequence against magnetic field $H$ to reproduce that in experiments in the wide range of model parameters of $K$ and magnetic anisotropy, as detailed in \cite{HayamiModel2}.

In Figure 4g-l, the spin textures for the phases I, II, III, IV, III' and V were calculated with ($H_{[100]}$, $H_{[001]}$, $K$) = (0, 0.1, 0.3), (0, 0.65, 0.3), (0, 1.0, 0.3), (1.0, 0, 0.2), (1.0, 0, 0.3) and (1.8, 0, 0.3), respectively. For Figure 4m-r, the topological charge density $n_{\rm sk} ({\bf r})$ was calculated following the procedure in \cite{HayamiModel2, SkNumberSep}.

\section*{Author contributions} S.S, T.A and Y.T supervised the project. N.D.K grew samples and measured the magnetic and transport properties with the assistance of R.T.  RXS measurements were carried out by T.N, S.G and N.D.K, with the assistance of K.K, Y.Y, H.S and H.N. S.H and Y.M performed Monte Carlo simulation. N.D.K, S.S, and S.H wrote the manuscript with inputs from co-authors. All the authors discussed the results and commented on the manuscript.

\section*{Acknowledgement} The authors thank R. Arita, K. Ishizaka, T. Hanaguri, Y. Yasui, C. J. Butler, T. Koretsune, T. Nomoto, M. Hirschberger, T. Kurumaji, Y. Ohigashi, K. Kolincio, L. Spitz and A. Kikkawa for enlightening discussions and experimental helps. RXS experiments were carried out under the approval of the Proposals nos 2018G570 and 2015S2-007 at the Institute of Material Structure Science, High Energy Accelerator Research Organization (KEK). This work was partly supported by Grants-In-Aid for Scientific Research (S) (grant no 21H04990), Grants-In-Aid for Scientific Research (A) (grant nos 18H03685, 20H00349 and 21H04440), Grants-In-Aid for Scientific Research (B) (grant no. 21H01037), Grant-in-Aid for Challenging Research (Exploratory) (grant no. 21K18595) and 	
Grant-in-Aid for Early-Career Scientists (grant nos 21K13873 and JP18K13488) from JSPS, PRESTO (grant nos JPMJPR18L5, JPMJPR20L8 and JPMJPR20B4) and CREST (grant no. JPMJCR1874) from JST, Katsu Research Encouragement Award of the University of Tokyo, Asahi Glass Foundation and Murata Science Foundation.

\newpage


\begin{thebibliography}{100}

\bibitem{SkXReviewFertTwo} A. Fert, N. Reyren, V. Cros, Nature Rev. Mater, {\bf 2017}, 2, 17031. 
\bibitem{SkXReviewTokura} N. Nagaosa, Y. Tokura, Nature Nanotech. {\bf 2013}, 8, 899.
\bibitem{SkXTheoryFirst} U. K. R\"{o}\ss ler,  A. N. Bogdanov, C. Pfleiderer, Nature {\bf 2006}, 422, 797.

\bibitem{MnSi} S. M\"{u}hlbauer, B. Binz, F. Jonietz, C. Pﬂeiderer, A. Rosch, A. Neubauer, R. Georgii, P. Boni, Science {\bf 2009}, 323 915 (2009).
\bibitem{TEMFeCoSi} X. Z. Yu, Y. Onose, N. Kanazawa, J. H. Park, J. H. Han, Y. Matsui, N. Nagaosa, Y. Tokura, Nature {\bf 2010}, 465, 901.

\bibitem{magnonic} M. Garst, J. Waizner, D. Grundler, J. Phys. D: Appl. Phys. {\bf 2017}, 50, 293002.

\bibitem{Cu2OSeO3_Seki} S. Seki, X. Z. Yu, S. Ishiwata, Y. Tokura, Science {\bf2012}, 336, 198.
\bibitem{CoZnMn_First} Y. Tokunaga, X. Z. Yu, J. S. White, H. M. Rønnow, D. Morikawa, Y. Taguchi, Y. Tokura , Nature Comm. {\bf 2015},6, 7638.
\bibitem{GaV4S8} I. K\'{e}zsm\'{a}rki, S. Bordács, P. Milde, E. Neuber, L. M. Eng, J. S. White, H. M. Rønnow, C. D. Dewhurst, M. Mochizuki, K. Yanai, H. Nakamura, D. Ehlers, V. Tsurkan, A. Loidl, Nature Mater. {\bf 2015}, 14, 1116 (2015).

\bibitem{Heusler} A. K. Nayak, V. Kumar, T. Ma, P. Werner, E. Pippel, R. Sahoo, F. Damay, U. K. Rößler, C. Felser, S. S. P. Parkin, Nature {\bf 2017}, 548, 561.

\bibitem{Gd2PdSi3} T. Kurumaji, T. Nakajima, M. Hirschberger, A. Kikkawa, Y. Yamasaki, H. Sagayama, H. Nakao, Y. Taguchi, T. Arima, Y. Tokura, Science {\bf 2019}, 365, 914.

\bibitem{GdKagome} M. Hirschberger, T. Nakajima, S. Gao, L. C. Peng, A. Kikkawa, T. Kurumaji, M. Kriener, Y. Yamasaki, H. Sagayama, H. Nakao, K. Ohishi, K. Kakurai, Y. Taguchi, X. Z. Yu, T. Arima, Y. Tokura , Nature Commun. {\bf 2019}, 10, 5831.


\bibitem{GdRu2Si2} N. D. Khanh, T. Nakajima, X. Z. Yu, S. Gao, K. Shibata, M. Hirschberger, Y. Yamasaki, H. Sagayama, H. Nakao, L. C Peng, K. Nakajima, R. Takagi, T. Arima, Y. Tokura, S. Seki, Nature Nanotech. {\bf 2020}, 15, 444.

\bibitem{Frustration1} T. Okubo, S. Chung, H. Kawamura, Phys. Rev. Lett. {\bf 2012}, 108, 017206.
\bibitem{Frustration2} A. O. Leonov, M. Mostovoy, Nature Commun. {\bf 2015}, 6, 8275.
\bibitem{AkagiJPSJ} Y. Akagi, Y. Motome, J. Phys. Soc. Jpn. {\bf2010}, 79, 083711.
\bibitem{AkagiPRL} Y. Akagi, M. Udagawa, Y. Motome, Phys. Rev. Lett. {\bf 2012}, 108, 096401. 
\bibitem{HayamiModel1} S. Hayami, R. Ozawa, Y. Motome, Phys. Rev. B {\bf 2017}, 95, 224424.
\bibitem{HayamiModel3} R. Ozawa, S. Hayami, Y. Motome, Phys. Rev. Lett. {\bf 2017}, 118, 147205.
\bibitem{BatistaModel1} Z. Wang, Y. Su, S. Z. Lin, C. D. Batista, Phys. Rev. Lett. {\bf 2020}, 124, 207201.

\bibitem{GdRu2Si2_STM} Y. Yasui, C. J. Butler, N. D. Khanh, S. Hayami, T. Nomoto, T. Hanaguri, Y. Motome, R. Arita, T. Arima, Y. Tokura,  S. Seki , Nature Commun. {\bf 2020}, 11, 5925.

\bibitem{Batista2} Z. Wang, Y. Su, S. Z. Lin, C. D. Batista, Phys. Rev. B {\bf 2021}, 103, 104408.
\bibitem{utesov} O. I. Utesov, Phys. Rev. B {\bf 2021}, 103, 064414.

\bibitem{RRuSi2} M. \'{S}laski, J. Mag. Mag. Mater. {\bf 1984}, 46, 114.
\bibitem{GdRu2Si2_PhaseDiagram} A. Garnier {\it et al}., J. Magn. Magn. Mater. {\bf 1995}, 140, 899.
\bibitem{RXS_Rule} M. Blume, in Resonant Anomalous X-Ray Scattering (eds Materlik, G., Sparks, C. J. and Fischer, K.), Elsevier, 1994.

\bibitem{HayamiModel2} S. Hayami, Y. Motome, Phys. Rev. B {\bf 2021}, 103, 024439.
\bibitem{Blugel} S. Heinze, K. von Bergmann, M. Menzel, J. Brede, A. Kubetzka, R. Wiesendanger, G. Bihlmayer, S. Blügel, Nature Phys. {\bf 2011}, 7, 713.




\bibitem{Meron1} X. Z. Yu, W. Koshibae, Y. Tokunaga, K. Shibata, Y. Taguchi, N. Nagaosa, Y. Tokura, Nature {\bf 2018}, 564,  95.

\bibitem{Meron3} S. Gao, H. D. Rosales, F. A. G. Albarracín, V. Tsurkan, G. Kaur, T. Fennell, P. Steffens, M. Boehm, P. Čermák, A. Schneidewind, E. Ressouche, D. C. Cabra, C. Rüegg, O. Zaharko, Nature {\bf 2020}, 586, 37.

\bibitem{Meron2} S. Z. Lin, A. Saxena, C. D.. Batista, Phys. Rev. B {\bf 2015}, 91, 224407.

\bibitem{THE}  A. Neubauer, C. Pfleiderer, B. Binz, A. Rosch, R. Ritz, P. G. Niklowitz, P. Böni, Phys. Rev. Lett. {\bf 2009}, 102, 186602.
\bibitem{EmergentEfield} T. Schulz, R. Ritz, A. Bauer, M. Halder, M. Wagner, C. Franz, C. Pfleiderer, K. Everschor, M. Garst, A. Rosch, Nature Phys. {\bf 2012}, 8, 301.

\bibitem{SkNumberSep} F. Eriksson, Math. Mag. {\bf 1990}, 63, 184.


\end{thebibliography}
\end{document}